\documentclass[letterpaper,pra,aps,floatfix,superscriptaddress,onecolumn,longbibliography,notitlepage
]{quantumarticle}
\pdfoutput=1
\usepackage[utf8]{inputenc}
\usepackage[english]{babel}
\usepackage[T1]{fontenc}
\usepackage{amsmath,amsthm,amsfonts,amssymb,xcolor}
\usepackage{graphicx}
\usepackage{microtype}
\usepackage{braket}
\usepackage[numbers,sort&compress,square]{natbib}

\usepackage{subcaption}

\makeatletter
\def\l@subsubsection#1#2{}
\makeatother

\usepackage[colorlinks=true,citecolor=blue,linkcolor=blue]{hyperref}
\usepackage[capitalize,nameinlink]{cleveref}

\theoremstyle{definition}

\newcommand{\RR}{{\mathbb{R}}}
\newcommand{\ZZ}{{\mathbb{Z}}}
\newcommand{\FF}{{\mathbb{F}}}

\newcommand{\stab}{\mathcal{S}}

\begin{document}
\title{Dynamically Generated Logical Qubits}

\author{Matthew B.~Hastings}                                                                                                                                                                                                                    \affiliation{Station Q, Microsoft Quantum, Santa Barbara, CA 93106-6105, USA}                                          \affiliation{Microsoft Quantum and Microsoft Research, Redmond, WA 98052, USA}                                                                                                                                                                                                                                                              

\author{Jeongwan Haah}  
\affiliation{Microsoft Quantum and Microsoft Research, Redmond, WA 98052, USA}                                                                                                                                                                                                                                                              
                                                                                                
\begin{abstract}
We present a quantum error correcting code with \emph{dynamically generated logical qubits}.
When viewed as a subsystem code, the code has no logical qubits.
Nevertheless, our measurement patterns generate logical qubits, 
allowing the code to act as a fault-tolerant quantum memory.
Our particular code gives a model very similar to the two-dimensional toric code,
but each measurement is a \emph{two}-qubit Pauli measurement.
\end{abstract}
\maketitle

A series of generalizations of quantum error correcting codes have been considered by many authors.
The simplest kind of code is a stabilizer code, where the checks are products of Pauli operators which are mutually commuting.  The toric code~\cite{Kitaev_2003} is a standard example of this code, and one of the most promising codes for practical realizations.  Implementing the toric code in two dimensions requires measurement of stabilizers which are products of at most {\it four} Pauli operators.
One can generalize stabilizer codes to subsystem codes~\cite{Poulin_2005}.  In this case, the operators measured need not be mutually commuting.  They generate a group that contains some central elements, called the stabilizers.  It is possible to implement the toric code as a subsystem code using measurements of checks which are products of at most {\it three} Pauli operators~\cite{Bravyi_2013}, and there is a subsystem color code~\cite{Bombin_2010} where checks are products of two Pauli operators.  One hope is that by reducing the number of qubits involved, the measurement circuit for the checks can be simplified.%
\footnote{
	Another subsystem code with two-qubit check operators 
	which is geometrically local in two-dimensions is the Bacon--Shor code~\cite{Bacon_2006}.  
	However, this code is not suitable for fault tolerance 
	without applying some further concatenation because the stabilizer weights scale with the number of qubits.
}
Another direction of generalization for the toric code is to regard the stabilizers of the toric code as terms in some quantum Hamiltonian, and instead consider Hamiltonians which are sums of noncommuting terms.  This direction leads to a wide range of models, including topologically ordered phases.

Here we propose a different direction of generalization.
We present the ``honeycomb code'' where checks are \emph{two}-qubit Pauli operators.
When regarded as a subsystem code, our code does not have any logical qubits.
Nevertheless, our particular measurement order still protects quantum information; 
indeed, on a torus it protects two logical qubits 
with a code distance proportional to the linear size of the torus.

The simplicity of the code may make it useful for applications, 
especially in architectures where the basic operation 
is a pairwise measurement such as Majorana devices~\cite{Karzig_2017}.
We will refer to such codes, where measurements are products of Paulis and the number of logical qubits is larger than the number that would be obtained when the code is viewed as a subsystem code, 
as codes with dynamically generated logical qubits, or ``Floquet codes.''

The subsystem color code with two-qubit check operators~\cite{Bombin_2010}
has some similarities and differences to our construction.
Both require a 3-colorable tiling of a two-dimensional plane 
to determine the check operators.
Both use periodic sequences of measurements.
However, 
in the subsystem color code 
any measurement schedule may be used without destroying the logical qubits, at least in the absence of errors,
since it has logical qubits in the usual sense.
In our honeycomb code, the measurement ordering is a crucial part of the construction;
 other measurement schedules may destroy the logical qubits.
In addition, we have a shorter measurement cycle and use fewer qubits at the given code distance.

We would like to distinguish our dynamical codes from the kinds of codes produced by 
random monitored measurements interspersed with rapidly scrambling unitaries~\cite{Fisher1,Ruhman}.
The scrambling induced by the unitary evolution 
acts as an error correcting code there to protect information against the measurements
and the combination of measurement and unitary can act as an encoder for an error correcting code~\cite{Gullans_2020}.%
\footnote{
  Though, due to the random nature of unitaries and measurements, it is not clear whether such a code would have a good decoder.
}
However, the models considered in that field are {\it not} suitable for fault tolerant applications,
since they will not protect against weak external noise introduced during the measurement and unitary process; 
at least, such correction seems not to have been considered and it does not happen in certain simple models (see \cref{app}).
The reason is that the measurements are not chosen in such a way that external noise will leave a strong enough fingerprint in the measurement outcomes.
Our model instead has the property that there are certain linear relations between measurement outcomes at different times in the absence of noise, and from these linear relations, noise can be detected and corrected.

In \cref{hc}, we define the honeycomb code.
Since the code has some unusual properties, 
in \cref{lc} we give a simpler one-dimensional code 
which we call the ``ladder code'' which can be analyzed more simply.
Finally, in \cref{sec:ec}, we show that error correction 
and even fault tolerance in the honeycomb code is possible.

\section{The Honeycomb Code}
\label{hc}

In this section, we introduce the honeycomb code, specified by
a time ordered sequence of check operators that are two-qubit Pauli operators.
If we ignore the time ordering and regard it as a subsystem code,
then the honeycomb code does not have any logical qubits.
Nonetheless, we can identify a subspace of dimension~$4$ (two qubits) at any moment,
and this subspace will be our logical qubits.
Naturally, we are led to ``instantaneous stabilizer groups'' and their dynamics.

\subsection{The Code}

We consider qubits arranged on vertices of a hexagonal (also called honeycomb)
lattice with periodic boundary conditions.
The edges are of three different types $x,y,z$.
At each vertex, three different types of edges meet.
The simplest choice is to pick $x,y,z$ to correspond to the three different directions of the edge.  
For each edge, we will define an operator called a ``check'' acting on the two qubits of the edge.
For an $x$ edge, the check is $XX$; for a $y$ edge, the check is $YY$, and for
a $z$ edge, the check is~$ZZ$.  Remark: in everything that follows, it suffices that each check be a product of two, possibly different, Paulis (for example, $XZ$) such that for every qubit, each check involving that qubit involves a distinct Pauli operator on that qubit.

\begin{figure}
\centering
\includegraphics[width=3in]{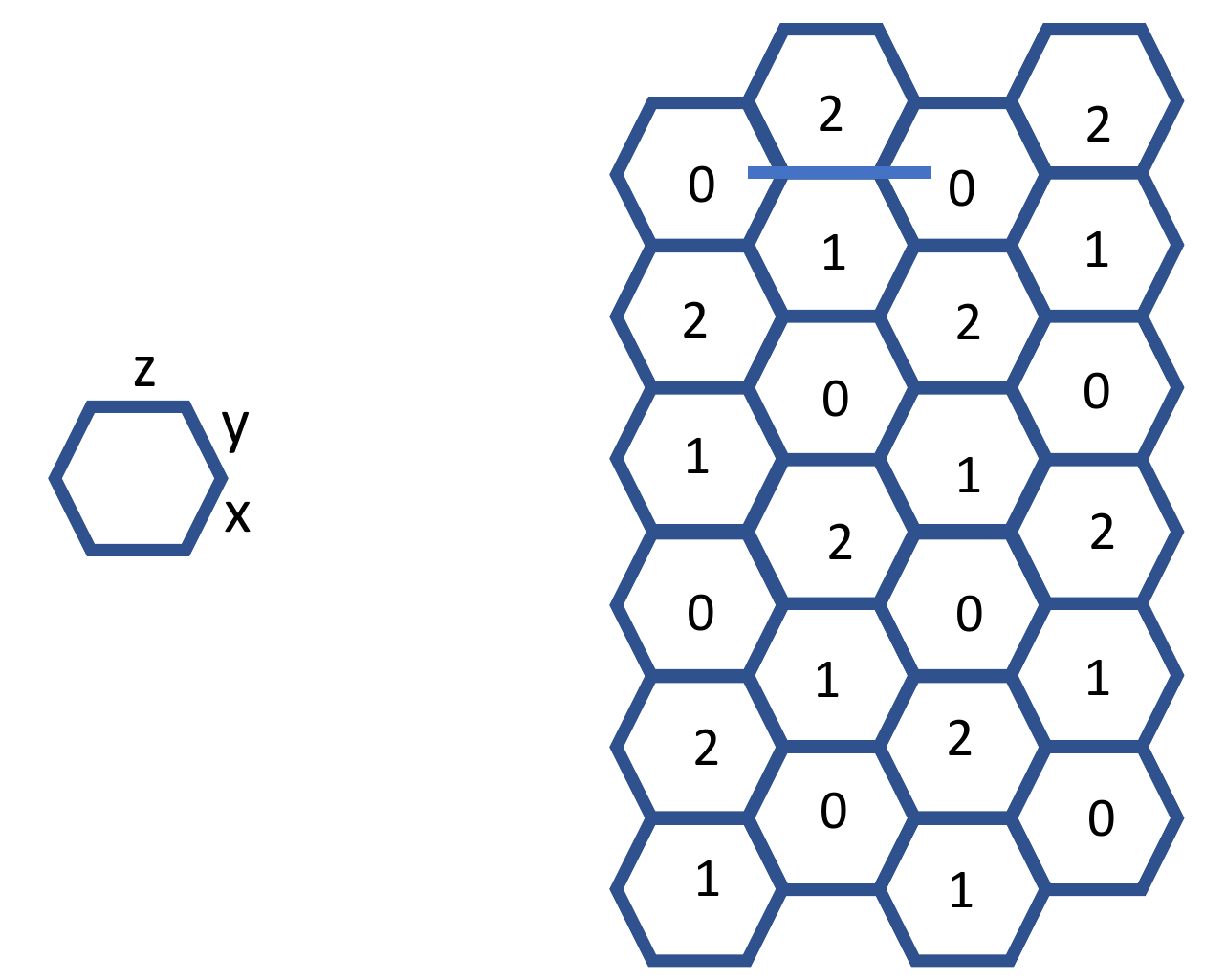}
\caption{
Labeling of plaquettes and edges.
Left side of figure shows three different types of edges, $x,y,z$ depending on direction.  Right side of figure shows hexagonal lattice.  The hexagons have been labeled by $0,1,2$ as described.  The slightly thicker longer line near the top connects two hexagons of type $0$, and hence the edge under it is a type $0$ edge.}
\label{fig:lat}
\end{figure}

The hexagons of the honeycomb lattice are 3-colorable,
meaning that we may label each hexagon by one of $0,1,2$
such that two neighboring hexagons have different labels.
Given this labelling of hexagons, 
we also label the edges by numbers $0,1,2$ by the rule that
every edge of label $a \in \{0,1,2\}$ connects 
two nearest hexagons of the same label $a$.
If an edge of type $a$ is slightly extended, its two endpoints would lie in hexagons of type $a$.
Note then that every edge is given two different labels, a letter $x,y,z$ and a number $0,1,2$,
so that there are $9$ types of edges.
See \cref{fig:lat}.

We measure the checks in distinct rounds, 
measuring first all checks labeled by $0$, 
then by $1$, then by $2$, repeating, 
so that in the $r$-th round we measure checks labeled by $r \mod 3$.
Note that every qubit participates in some check at every round.

This completes the definition of our honeycomb code.
Our definition is perhaps unusual 
since we have not identified a fixed subspace of the full Hilbert space of the qubits, 
in which one would encode logical qubits.
However, a useful subspace is dynamic as we will show.

\subsection{Properties as Subsystem Code}

A subsystem code is defined as follows.  See \cite{Poulin_2005} for the original reference, and \cite{Bombin_2010} contains a useful review.  There are a set of operators (which correspond to measurements used in detecting and correcting errors) called checks.  These checks are products of Pauli operators on qubits.
The checks generate a group called the ``gauge group.''
The center of the gauge group is called the ``stabilizer group.'' 
The gauge group can then be generated by the stabilizer group and by some other group which is isomorphic to a tensor product of Pauli groups; this other group is regarded as acting on certain ``gauge qubits.''
 Nontrivial logical operators correspond to operators which commute with the gauge group but which are not in the stabilizer group.
 
 At this point, a remark on terminology is necessary.  The term ``gauge'' is highly overloaded, being used as above to describe a certain group in a subsystem code but also (and this will be useful for us later) to describe a ``gauge field'' which is a certain operator defined in a Majorana representation; see for example \S2.3 of~\cite{Kitaev_2006} though we will explicitly define what we mean by gauge fields in \cref{majrep}.  Thus, when we use the terms ``gauge group'' and ``gauge qubit,'' these will refer to the terms in the above paragraph.  We will continue to call the operators that we measure ``checks,'' rather than ``gauge operators'' as they are sometimes called.
 We reserve the term ``gauge field'' for the operators defined in \cref{majrep}.
 
 Consider the honeycomb code on a torus with $n_p$ hexagonal plaquettes.  Then, there are are $2n_p$ qubits.
 There are $3n_p$ edges; however, there is a redundancy of the checks since the product of all checks is the identity.  Hence, the gauge group has dimension $3n_p-1$.  
 
The product of checks on any {\it cycle}\footnote{A $1$-chain is an assignment of $0$ or $1$ to each edge of the lattice, and a $1$-cycle (or simply a cycle for short) is a $1$-chain whose boundary vanishes mod $2$.  Equivalently,
a $1$-chain is a cycle if, for every vertex, the number of edges
incident to that vertex which are assigned a $1$ is even.  By ``product of checks on a cycle,'' we mean the product over all edges with a $1$, of the check on that edge.  The ordering of the product may be chosen arbitrarily, as the result is the same up to a sign.} on the lattice is a stabilizer, and indeed these are all the stabilizers.  The stabilizers corresponding to homologically trivial paths are generated by paths on plaquettes, and we call the corresponding operators ``plaquette stabilizers.''  However, there is again a redundancy, as the product of all those plaquette stablizer is the identity.  The stabilizer group is generated by these plaquette stabilizers as well as by stabilizers for two homologically nontrivial cycles wrapping different directions of the torus.  Thus, the stabilizer group has dimension $n_p+1$.

Hence, there are $g=((3n_p-1)-(n_p+1))/2=n_p-1$ gauge qubits.  Since the stabilizer group has dimension $s=n_p+1$ and $g+s$ is equal to the number of qubits, there are no logical operators {\it when this code is regarded as a subsystem code}.

\subsection{Instantaneous Stabilizer Groups}

Recall that given a state stabilized by a Pauli stabilizer group $\stab$,
a measurement by a Pauli operator~$P$ projects the state to another Pauli stabilizer state,
whose stabilizer group can be tracked as follows.
\begin{itemize}
  \item[(a)] If $P \in \stab$ or $-P \in \stab$, 
  	then the measurement of $P$ is deterministic and reveals $P = \pm 1$
    and the stabilizer group remains the same.
  \item[(b)] If $P \notin \stab$ and $-P \notin \stab$, but if $P$ commutes with every element of $\stab$,
    then the post-measurement stabilizer group $\stab'$ is increased from $\stab$ to include $\pm P$
    ($\stab' = \langle \stab, \pm P \rangle$),
    where the sign is given by the measurement outcome.
    When the underlying state is maximally mixed within the stabilized subspace,
    the two outcomes $\pm 1$ are equally probable.
  \item[(c)] If $P \notin \stab$ and if $P$ anticommutes with some element $Q \in \stab$,
    then the post-measurement stabilizer group does change but the size remains the same.
    Specifically, if $\stab = \langle \stab_0 , Q \rangle$ where $\stab_0$ is a subgroup of $\stab$
    consisting of those that commute with $P$,
    then the post-measurement stabilizer group is $\stab' = \langle \stab_0 , \pm P \rangle$
    where the sign is given by the measurement outcome.
    The two outcomes are always equally probable.
\end{itemize}

Suppose we start in a maximally mixed state, which is a stabilizer state of the trivial stabilizer group,
and begin measuring checks in the pattern above,
starting with round $0$.
Here we assume that every measurement is noiseless;
we will address effects of noise in \cref{sec:ec}.
After any number of rounds, the state is a Pauli stabilizer state,
specified by an ``instantaneous stabilizer group (ISG).''
Let us identify the ISG after each round of the honeycomb code.
We will not consider signs of the stabilizers here in this section,
though the signs will be important in \cref{sec:ec}.
It is important to note that the product of the checks
over all edges of any closed loop commutes with any check. 
In particular, each hexagon supports a weight~6 operator, which we call a plaquette stabilizer,
that commutes with every check.

The key observation to understand what follows is that measuring checks in rounds $r-1$ and $r$ 
will infer the plaquette stabilizers on plaquettes of type $r+1 \mod 3$.

\begin{enumerate}
  \item The ISG is initially trivial (containing only the identity operator).
  \item After round~$0$, it is clear that the ISG $\stab(0)$ is generated by the checks of type $0$.
  \item After round~$1$, the ISG $\stab(1)$ is generated by the checks of type $1$ and the plaquette stabilizers on hexagons of type $2$.
  To see this, note that the six qubits of a type~$2$ hexagon has not interacted with any other type~$2$ hexagon. Hence, it suffices to verify the claim for one hexagon of type~$2$.
  We add three checks to $\stab(0)$ one by one, and the claim follows by applying (c) above twice and (b) once.
  \item After round~$2$, the ISG $\stab(2)$ is generated by the checks of type~$2$ and the plaquette stabilizers of type~$2$ and type~$0$.
  The reason is similar: Since the plaquette stabilizers of $\stab(1)$ commute with the type~$2$ checks,
  we may consider the subgroup of $\stab(1)$ generated by the type~$1$ checks.
  The situation is the same as in the transition $\stab(0) \to \stab(1)$.
  \item \emph{On subsequent rounds $r\geq 3$, the ISG $\stab(r)$ is generated by checks of type $r \bmod 3$ and all the plaquette stabilizers.}
  The proof is by induction: $\stab(2)$ contains all the plaquette stabilizers of type~$2$ and~$0$,
  and the type~$2$ checks of $\stab(2)$ and the type~$3$ checks at round~$3$ generates
  type~$1$ plaquette stabilizers and type~$3$ checks. For round $r > 3$, we know $\stab(r-1)$ contains
  all plaquette stabilizers. The type~$(r-1\bmod 3)$ checks in $\stab(r-1)$ and the type~$(r\bmod 3)$ checks at round $r$,
  replace the type~$(r-1\bmod 3)$ checks with the type~$(r\bmod 3)$ checks.
\end{enumerate}

It is crucial that for any $r$, the ISG $\stab(r)$ 
never contains homologically nontrivial ``long'' loop operators.
We have remarked that the product of all the checks along a closed loop commutes with every check.
Here, the loop can be a homologically nontrivial loop
which is not the circumference of any combination of plaquettes.
This long loop operator belongs to the center of the ``gauge group'' of the subsystem code,
but our specific measurement dynamics keeps the long loop operators away 
from the instantaneous stabilizer group.

We will mainly consider the ISG for $r\geq 3$, 
when the ISG reaches a steady state depending only on $r \bmod 3$.
If there are $n_p$ plaquettes on a torus, 
there are $n_p$ checks of types $r \bmod 3$, so we have given $2n_p$ generators for $\stab(r)$.
However, these generators are not independent:
the product of all plaquette stabilizers is the identity and also the product of check of type $r\mod 3$ with plaquette stabilizers of type $r\bmod 3$ is the identity.
So, the ISG has dimension $2n_p-2$, and hence there is a $2^2 =4$ dimensional subspace
stabilized by ISG at any moment.
Thus, our choice of measurement sequence is important.
If, for example, we had instead chosen to measure operators of types $x,y,z$ in rounds $r=0,1,2 \mod 3$ respectively, 
then the long loop operators would have been in the instantaneous stabilizer group,
destroying the logical subspace.

In fact, the code described by the instantaneous stabilizer group is a toric code,
up to a bounded-depth quantum circuit.
Even disregarding error-correction properties of the honeycomb code,
this may be a useful way to rapidly prepare toric code states 
using only $4$ rounds of pairwise measurements ($r=0,1,2,3$).

\subsection{Logical operators and embedded toric code}

\paragraph{Inner and outer logical operators.}

With two logical qubits, we need to find a basis of four logical operators.
One type of logical operators is (up to the instantaneous stabilizer group)
the product of check operators on a homologically nontrivial cycle.  
This gives us two logical operators.
We call these ``inner logical operators'' since they belong to the stabilizer group as a subsystem code.
The other type of logical operator is shown on \cref{fig:dynsuper} below.
We call these ``outer logical operators'' since they do not belong to the stabilizer group as a subsystem code.
Note that an outer logical operator is specific to a given round
and carries a sign that depends on the history of check outcomes;
we will elaborate on this below.
This gives us two more logical operators on a torus.
The outer and inner logical operators act as logical $X,Y$ operators on the two qubits, respectively.
Of course, it is completely arbitrary which logical operator corresponds to logical $X$ or $Y$.

We can distinguish inner and outer logical operators by their partial implementations.
The outer logical operator has the property that 
if we terminate the logical operator to a string with endpoints, 
then at the endpoints it anticommutes with some plaquette stabilizers.
The plaquette stabilizers form a static subgroup of ISG, 
which is the intersection of ISG over all rounds $r \ge 3$.
(This static subgroup of ISG consists of all homologically trivial elements 
of the stabilizer group of the honeycomb code, regarded as a subsystem code.)
The inner logical operator however can be terminated to a string with endpoints 
in a way that it commutes with all plaquette stabilizers, 
as we simply may take the product of gauge operators along an open path.
Similar to the outer logical operators,
an inner logical operator corresponds to a nontrivial homology cycle
consisting of edges of the honeycomb lattice,
and two inner logical operators of the same homology class are equivalent up to ISG.

Perhaps a more important distinction between inner and outer logical operators is their dynamics.
The inner logical operator commutes with all checks,
and hence if we initialize the logical qubits in an eigenstate of an inner logical operator,
and later measure the inner logical operator,
then we will retrieve the same eigenvalue.
This means that the inner logical operator is independent of the measurement round $r$,
even though the ISG is constantly changing.
In contrast, the dynamics of the outer logical operators is nontrivial.
To understand this dynamics we make a detour.

\paragraph{Embedded toric code.}

\begin{figure}
\centering
\includegraphics[width=2in]{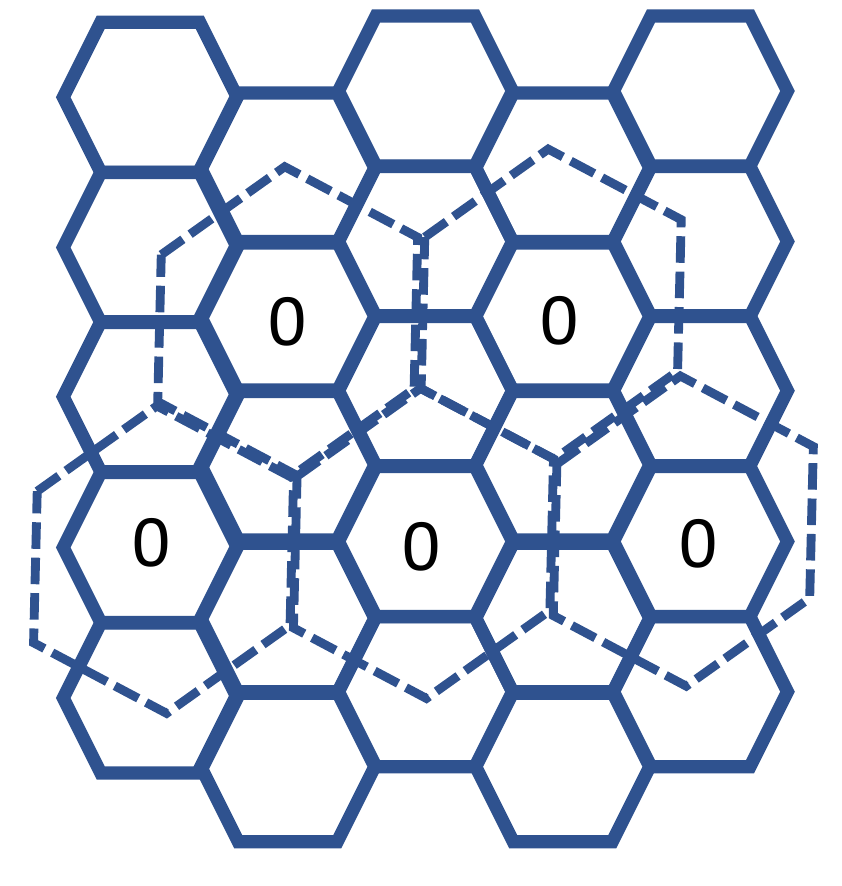}
\caption{
ISG corresponds to toric code after disentangling.
After round $r = 0 \bmod 3$, each type~$0$ check turns the two qubits in its support into one effective qubit.
Hence, after disentangling one qubit for each type~$0$ edge, 
the remaining qubits have the state of the toric code on a hexagonal superlattice.
}
\label{fig:toric}
\end{figure}

To examine the nontrivial dynamics of outer logical operators, 
it will be useful to map the ISG of the honeycomb code onto that of the toric code.
Each check of a given type involves two qubits.
Since the measurement of a check projects the state of the two qubits down to a two-dimensional subspace,
we may regard the pair of qubits in the support of a check as one effective qubit.  
More concretely, a local quantum circuit
consisting of one CNOT for each $x$ or $z$ edge and one control-$Y$ for each $y$ edge,
maps all checks of type~$r$ to single-qubit Paulis,
which disentangles a qubit out for each edge.

The ISG is generated by those checks as well as the plaquette stabilizers.
We claim that plaquette stabilizers are the stabilizers of a toric code on a hexagonal superlattice, after this disentangling.
A picture makes it clear; see \cref{fig:toric}.
Assuming measurement round $r=0 \bmod 3$ is just completed, 
we have drawn lines (shown dashed in the figure) across the type $0$ edges.
After disentangling, there is one qubit per such edge, so there is one qubit for each dashed line segment.
That is, the disentangling circuit turns the hexagonal lattice with qubits on vertices 
into a hexagonal superlattice with qubits on edges.
Drawing the dashed lines as they are, 
we see that each type~$0$ plaquette stabilizer acts on the six qubits corresponding to dashed lines 
on the dashed superplaquette surrounding the given original plaquette, 
while each type~$1$ or~$2$ plaquette stabilizer acts on the three qubits corresponding to dashed lines 
terminating at the center of the given original plaquette.
Indeed, one may verify that the type~$0$ plaquette stabilizers correspond to 
plaquette stabilizers of a hexagonal lattice toric code 
while the type~$1$ or~$2$ plaquette stabilizers correspond to vertex stabilizers.

\begin{figure}
	\centering
	\includegraphics[width=0.8\textwidth]{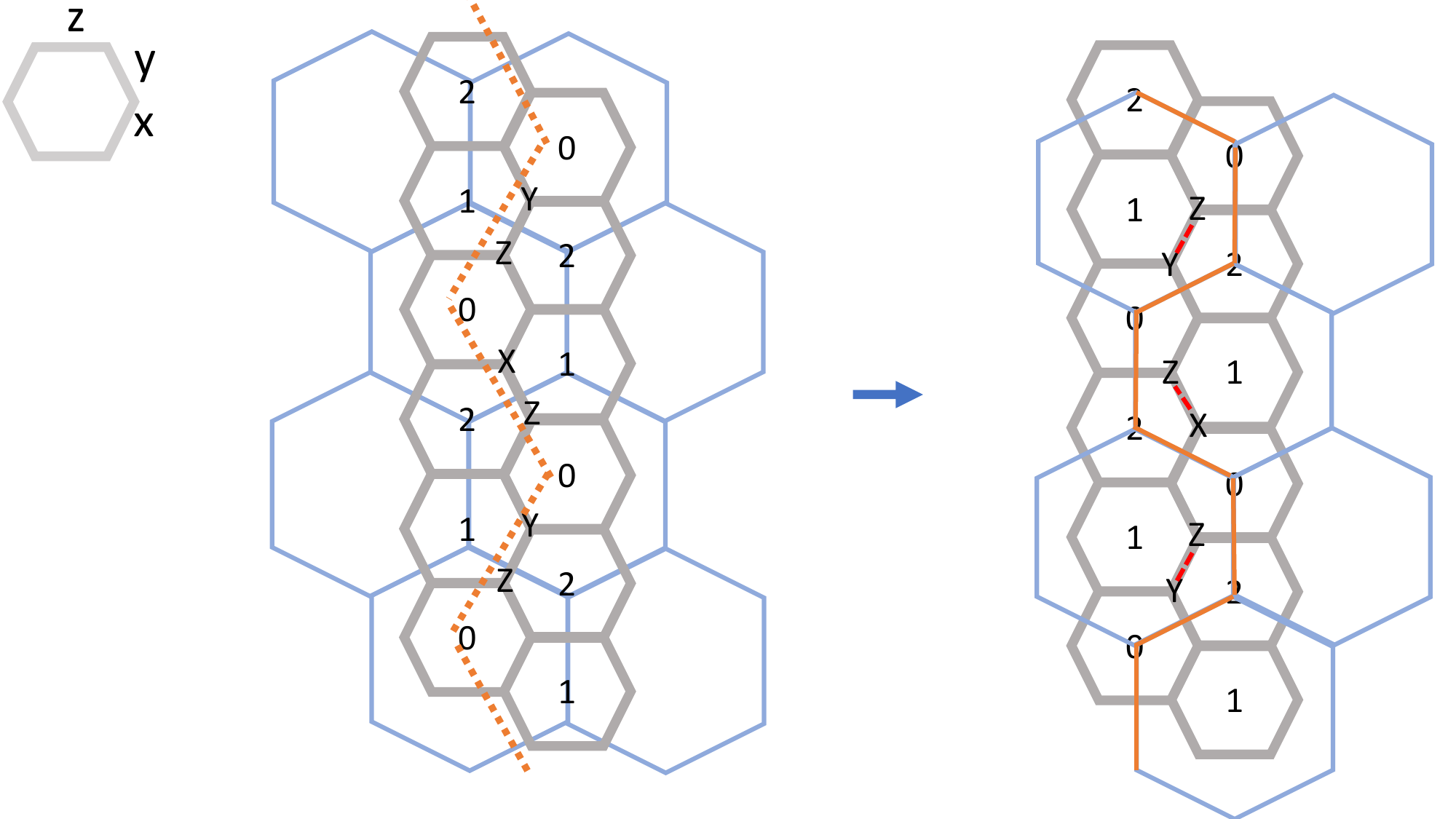}
	\caption{
		Dynamics of our honeycomb code as code deformation.
		The left figure shows an outer logical operator~$Q(r)$
		that is a magnetic operator of the toric code on the hexagonal superlattice, 
		where it is assumed that round~$r = 0 \bmod 3$ is just completed. 
		The right figure shows another outer logical operator $Q(r+1)$ that is an electric operator of the toric code,
		after round~$r+1$.
		The operator~$Q(r+1)$ is chosen such that the difference $Q(r+1)Q(r)^\dagger$ is an element of the ISG~$\stab(r)$ at round~$r$.
		Therefore, this shows how an outer logical operator evolves under our measurement dynamics.
		Since the transition ``magnetic $\leftrightarrow$ electric'' occurs for an odd number of times per period,
		every outer logical operator gets multiplied by a parallel inner logical operator every period,
		up to signs.
	}
	\label{fig:dynsuper}
\end{figure}

The toric code on the hexagonal superlattice has string and dual string operators,
which can be written in terms of operators on our honeycomb code.
When a dual string in the superlattice (the orange dashed line on the left in \cref{fig:dynsuper})
crosses an edge with check operator $a \otimes a$ ($a \in \{ X,Y,Z\}$) in the superlattice 
we write a two-qubit operator $a' \otimes a''$
where $a,a',a''$ are all distinct;
for example, after round $r = 0 \bmod 3$, on a type~$0X$ edge we write $Z\otimes Y$.
(We could write $Y\otimes Z = (X\otimes X)(Z\otimes Y)$, which is equally good.)
The constructed operator along the dual string on the superlattice is the magnetic operator of the toric code,
one that transports a superplaquette excitation $m$.
Similarly, when a direct string (rather than dual string)
passes an edge in the superlattice we write a one-qubit operator
that is a half of the check on that edge;
it does not matter which half we choose.
The orange solid line on the right in \cref{fig:dynsuper} depicts an example after round $r+1 = 1 \bmod 3$.
The constructed operator along the direct string on the superlattice is the electric operator of the toric code,
one that transports a supervertex excitation $e$.
These constructions give \emph{outer} logical operators of our honeycomb code
since a partial implementation will violate some stabilizer of the toric code,
which is a plaquette stabilizer of the honeycomb code.

\paragraph{Dynamics.}

The representative~$Q(r)$ chosen on the left of \cref{fig:dynsuper}
anticommutes with some element of the ISG~$\stab(r+1)$ at round~$r+1$.
However, there is an representative~$Q(r+1)$ that commutes with all elements of~$\stab(r+1)$ and~$\stab(r)$,
which is drawn on the right of \cref{fig:dynsuper}.
Observe that with respect to~$\stab(r)$ the operator~$Q(r)$ is magnetic,
while with respect to~$\stab(r+1)$ the modified operator~$Q(r+1)$ is electric.
We must ask how canonical this transition $Q(r) \to Q(r+1)$ is.
The operator $Q(r)$ could be $Q'(r)$ where the difference $Q'(r)Q(r)^\dagger$ is an element of $\stab(r)$ 
but such a difference cannot change the fact that $Q'(r)$ is magnetic.
Similarly, at round $r+1$, the operator $Q(r+1)$ is equivalent to any other operator $Q'(r+1)$
where the difference $Q'(r+1) Q(r+1)^\dagger$ is an element of $\stab(r+1)$
and it remains invariant that $Q'(r+1)$ is electric.
Hence, as long as we fix the embedding at each measurement round 
from the toric code into the hexagonal superlattice on our honeycomb lattice,
the conclusion that a magnetic operator transitions to an electric operator remains valid.
It is left to the reader to verify that any electric operator at round $r$ goes to a magnetic operator at round $r+1$.%
\footnote{
If we considered triangular superlattices instead of hexagonal superlattices,
what we called electric would be magnetic and vice versa.
We only consider hexagonal superlattices throughout.
}

One may regard the dynamics from round~$r$ to~$r+1$ as quantum state transfer.
After round~$r$ one half of the degrees of freedom are in a product state due to the checks,
and the other half forms a toric code state.
After round~$r+1$ the state is pushed to a different corner of the Hilbert space by projections (checks at round~$r+1$)
to form another toric code state,
while mapping logical $X$ operators (magnetic) to logical $Z$ operators (electric) and vice versa.

Up to signs, the ISG $\stab(r)$ is periodic in $r$ with period~$3$,
and there are three superlattices associated with this periodicity.
Since the period is an odd number, 
after one period (three measurement rounds) every electric operator on $\stab(r)$
goes to a parallel magnetic operator on $\stab(r+3)$ and vice versa.
Since $\stab(r)$ and $\stab(r+3)$ are the same groups up to signs,
this transition is meaningful as an automorphism on the stabilizer code defined by $\stab(r)$.
But, the difference between an electric and a parallel magnetic operator is an operator transporting a fermion,
so the difference cannot be an outer operator;
we have found four outer logical operators and they are transporting bosons $e$ and $m$.
Therefore, the difference must be an inner logical operator.
That is, every outer logical operator gets multiplied by a parallel inner logical operator every period.
We conclude that the dynamics of outer logical operators has period~$6$
even though the measurement sequence has period~$3$ if we ignore the signs.

Note that the signs are important to use our honeycomb code.
Microscopically, the transition $Q(r) \to Q(r+3)$ of an outer logical operator
is performed by a product of check operators at rounds~$r,r+1$, and $r+2$.
The measured signs of individual check operators are completely random,
and hence we must record these signs to correctly infer the eigenvalue of $Q(r+3)$.
In general, if we keep the honeycomb code state for many rounds,
an outer logical operator carries a sign that depends on the checks on the membrane in $2+1$-dimensional spacetime
that hangs from the present outer logical operator and extends to the past up to initialization step.
This is unusual in view of conventional stabilizer codes
where all stabilizer elements have signs that are fixed once and for all.

\begin{figure}
\centering
\includegraphics[width=2in]{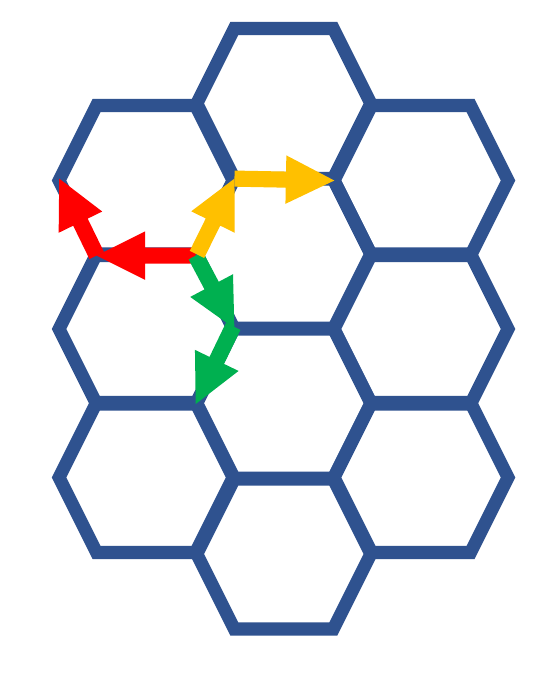}
\caption{
An example of three inner operator segments to show that it creates a fermion,
drawn in three different colors with arrows.
The center point is $0$, while the other ends of the inner operators at $a,b,c$.
Their commutation relation shows that an inner operator segment transports a fermion.
}
\label{fig:fermion}
\end{figure}

\paragraph{Fermions.}
The conclusion that the inner logical operator transports a fermion
can also be understood via a ``T-exchange'' process 
(see for example~\cite[Fig.1]{Kawagoe_2020}).
Pick four points on the lattice, called $a,b,c$ and $0$.
Draw three different inner operators $O_a, O_b, O_c$, each of which is a product of check operators: 
they start at $0$ and end at $a,b,c$, respectively.
These operators happen to be unitaries.
We show an example in \cref{fig:fermion}.
The product $O_u^\dagger O_v$ creates an anyon pair at $u\neq v$, moving a particle from $u$ to $v$,
where $u,v = a,b,c$.
The product $(O_c^\dagger O_b) (O_a^\dagger O_c) (O_b^\dagger O_a)$ corresponds to exchange of two anyons
and is equal to $-1$ as may be seen by a direct computation:
all check operators in $O_a,O_b,O_c$ other than the checks which include $0$ trivially cancel in this product 
and the product is equal to (up to an arbitrary relabeling of Pauli operators) $XYZXYZ = -1$.
This shows that the inner operator transports a fermion.

Indeed, it should be no surprise that the inner operator is a fermion: 
the multiplication of the outer operator by a fermion after each round of measurements then 
means we interchange $e \leftrightarrow m$.
Both particles are bosons.
If instead the inner particle were a boson, then we would have
interchange boson and fermion after a given period of measurements!

\subsection{Majorana Representation}
\label{majrep}

The honeycomb code has a convenient representation in terms of Majorana fermions.  
Most of our results in this paper do not need the Majorana representation, so readers unfamiliar with this representation
may skip this subsection.

The mathematical representation was used in \cite{Kitaev_2006} to analyze the Kitaev honeycomb model.  The Kitaev honeycomb model and the honeycomb code are related: the honeycomb model has a Hamiltonian which is the sum of checks in the code.

In this Majorana representation, we represent each qubit by four Majorana operators, denoted $\gamma^0_j,\gamma^X_j,\gamma^Y_j,\gamma^Z_j$, where $j$ labels the qubit.  Then, the Pauli operators $X_j,Y_j,Z_j$ are represented by $i\gamma^0_j\gamma^X_j,i \gamma^0_j\gamma^Y_j, i\gamma^0_j\gamma^Z_j$, respectively.  The Majoranas are subject to the requirement $\gamma^0_j \gamma^X_j \gamma^Y_j \gamma^Z_j=1$ for all $j$; we call this requirement a ``gauge constraint.''

This representation also describes one possible physical realization of the code, using so-called ``tetrons''~\cite{Karzig_2017}, where each qubit really is represented by four Majorana modes, subject to the gauge constraint due to charging energy.  This realization may be particularly suitable for the honeycomb code, since the measurement of the check operators is the native operation in that realization (i.e., Clifford gates in that realization are built by measuring products of Paulis, so it is simpler to measure such a product than it is to implement a CNOT gate).

For each edge $(j,k)$ between a pair of vertices $j,k$, we define an operator $t_{jk}=\gamma^a_j \gamma^a_k$, where $a$ is $X,Y,Z$ depending on whether the edge is type $x,y,z$.
We may regard these $t_{jk}$ as ``gauge fields.''  The product of $t_{jk}$ around any closed path is equal (up to a scalar) to the product of checks along that path.
The operators $t_{jk}$ commute with all checks (when the checks are written in terms of Majoranas), although the $t_{jk}$ do not commute with the gauge constraint.

The physics of the honeycomb code is then clear.
Suppose we ignore the gauge constraint.  Then, the $t_{jk}$ can be taken as scalars, with their product around any plaquette equal to $+1$.
The other fermions $\gamma^0_j$ are driven by the checks between different states with zero correlation length, i.e., states where after measuring checks of type $a$ for $a\in\{0,1,2\}$, the products $i\gamma^0_j \gamma^0_k$ have expectation value $\pm 1$ whenever $(j,k)$ is of type $a$.
We can then take this state and impose gauge invariance by projecting onto states obeying the gauge constraint.%
\footnote{
	In \cite{Kitaev_2006}, so-called $A$ and $B$ phases were studied.  
	The $A$ phases correspond to topologically trivial gapped phases of a quadratic Hamiltonian for~$\gamma^0$.  
	In a limiting case (the corners of Fig.5 of \cite{Kitaev_2006}), 
	the $A_x,A_y,A_z$ phases have zero correlation length 
	and $i\gamma^0_j \gamma^0_k$ has expectation value~$\pm 1$ 
	whenever $(j,k)$ is of type $x,y,z$, respectively, 
	while for us the expectation is~$\pm 1$ when $(j,k)$ has type~$a$ for $a\in\{0,1,2\}$.
}

The inner logical operators are products of the gauge fields around homologically nontrivial cycles.
Each outer logical operator should anticommute with some inner logical operator, while commuting with the plaquette stabilizers, the gauge constraint, and the checks on edges of type $r\mod 3$ after round $r$.

If it were not for the gauge constraint, such an outer logical operator would be easy to write down: draw any cycle on the dual lattice.  Then, take the product over
edges $(j,k)$ cut by that cycle of an operator $\gamma^a_j$ where $a\in {X,Y,Z}$ depending on whether the edge is type $x,y,z$.  Equivalently, one could take operator $\gamma^a_k$ on such an edge.
However, this attempt at an outer logical operator may not commute with the gauge constraint
and with the checks on edges.
To solve this problem, after round $r$, we may try multiplying the operator by products $\gamma^0_l \gamma^0_m$ for edges $(l,m)$ of type $r \mod 3$.
An appropriate choice of such operators to multiply by gives the outer logicals of the last section. 

\section{Ladder Code}
\label{lc}

Perhaps the most surprising property of the honeycomb code is that fault tolerance is possible.  How can one correct errors, since we can terminate an inner logical operator in such a way that the endpoints commute with all plaquette stabilizers?  How can such errors be detected?
To better understand how this can work, before giving the proof in the next section, here we will consider a simple ladder model which also has dynamically generated logical qubits.

\subsection{Code definition}

\begin{figure}
  \centering
  \includegraphics[width=0.8\textwidth]{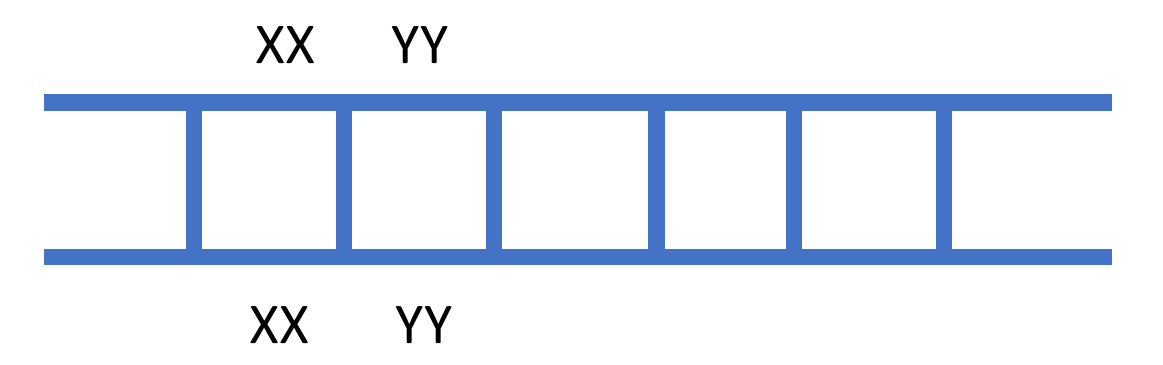}
  \caption{Ladder code.  There is one qubit per vertex.  Checks on vertical legs are $ZZ$ on the two qubits.  Checks on horizontal legs are alternately $XX$ or $YY$; some of the horizontal checks are shown.
  }
  \label{fig:ladder}
\end{figure}

The code is as shown in \cref{fig:ladder}.  There are two legs of a ladder.  The ladder is ``on its side,'' so that rungs of the ladder go vertically and legs go horizontally.  There is one qubit on each vertex.  Vertical checks are all $ZZ$.  Horizontal checks alternate $XX$ and $YY$ on each leg, as shown.  An $XX$ check on one leg is directly above an $XX$ check on the other leg, as shown.
The ladder is periodic, with an even number of rungs, so that left and right ends are attached.

Instead of using a three round repeating pattern to measure checks, we use a four round pattern.  We measure vertical $ZZ$ checks, then horizontal $XX$ checks, then vertical $ZZ$ checks again, then horizontal $YY$ checks, in rounds $r=0,1,2,3 \mod 4$, respectively.
The reason for using this four round repeating pattern is that if we instead had a three round repeating pattern (such as $ZZ$ checks, then $XX$ checks, then $YY$ checks, on two successive rounds we would measure all $XX$ checks, then all $YY$ checks, and so we would measure an inner logical operator, i.e. we would measure the product of all horizontal checks on a single leg of the ladder.

The ISG is easy to describe for $r\geq 4$.  The ISG is generated by the plaquette stabilizers, which are products of checks around a {\it square} of the ladder, and also the most recently measured checks.

A representative of the inner logical operator is the product of checks on a leg of the ladder, e.g., the product of Pauli $Z$ over the bottom leg.
An outer logical operator which anticommutes with this inner logical operator is the product $XX$ on any given vertical rung after $r=1 \mod 4$ and the product $YY$ after $r=3\mod 4$.  After $r=0,2 \mod 4$, we may take either the product $XX$ or the product $YY$ on a rung as the outer logical operator: they differ by the product $ZZ$ which is the check that was just measured on that rung.

\subsection{Fault tolerance}

The input for a decoder consists of certain linear combinations of measurement outcomes of the checks,
called ``syndrome'' bits.
They are designed to assume a trivial value deterministically in the absence of any faults 
and to provide sufficient information to infer likely faults.
Our syndrome bits are simply the plaquette stabilizers' eigenvalues,
but recorded \emph{every time} they can be inferred from recent measurement outcomes.
Since the plaquette stabilizers are always present in the ISG at any moment,
we know that they must assume $+1$ in the absence of any noise.

Concretely, suppose we have just finished $YY$ measurement in round $-1 \bmod 4$.
In the next round, $r = 0 \bmod 4$, we measure $ZZ$. 
The product of outcomes of $YY$ from the previous round and those of $ZZ$ from the current round,
gives the eigenvalues of the plaquettes that straddle $YY$ links.
Next ($r = 1 \bmod 4$), we measure $XX$ whose outcomes are combined with those of $ZZ$
to give the eigenvalues of the plaquettes that straddle $XX$ links.
Next ($r = 2 \bmod 4$), we measure $ZZ$ whose outcomes are combined with those of $XX$
to give the eigenvalues of the same set of plaquettes.
That is, we record the eigenvalues of the plaquettes over $XX$ links, twice in a row.
Next ($r = 3 \bmod 4$), we measure $YY$ whose outcomes are combined with those of $ZZ$
to give the eigenvalues of the plaquettes over $YY$ links.
Thus, we have completed one period of our measurement sequence,
and we have recorded the eigenvalues of all plaquettes, twice for each.
They are our syndrome bits.

Now we examine how we can use the syndrome bits.
A single-qubit Pauli error at any point will flip two of the checks,
so (the eigenvalue of) at least one plaquette will be flipped.
This is no different from a usual situation with stabilizer codes
where a single-qubit Pauli error anticommutes with some stabilizer.
Such a violation of a plaquette stabilizer persists indefinitely.
Since the instantaneous stabilizer code of the ladder code has code distance~$2$,
we conclude that the ladder code is error detecting with effective distance~$2$.

More interestingly, 
inasmuch as the classical repetition code of quantum distance~$1$ can correct many classical bit flip errors,
the ladder code can \emph{correct} certain type of errors.
Consider a chain of ``check errors,'' each of which is by definition a unitary operator equal to some check operator.
The chain of all check errors along the bottom leg of the ladder is an inner logical operator,
and we wish to correct any chain of check errors of sufficiently small weight.
For simplicity, we only consider $XX$ and $YY$ check errors on the bottom leg.
Suppose an $XX$ check error occurs right after the $YY$ measurement round.
The next round is to measure $ZZ$, two of which will be flipped by the error,
which in turn flips two plaquettes (two syndrome bits) over $YY$ links.
After $ZZ$ are measured, the next is to measure $XX$, into which the error is absorbed.
After one period of the measurement sequence,
the net effect of the $XX$ check error is to flip two syndrome bits,
which exist only in one time slice.
Unlike a single-qubit Pauli error, this check error is short-lived.
If we have a chain of consecutive check errors right after the round of $YY$ measurement at round, say, $-1$,
then exactly two plaquette stabilizers over $YY$ links, which are inferred after round $0$, are flipped,
and they are at the ends of the error chain.

Above, we have inserted check errors only at a particular time slice,
but a similar argument applies for any time slice.
Hence, in a low error rate regime,
if we knew that errors are check errors only,
we can correct errors confidently by a minimum-weight matching of nontrivial syndrome bits.
Note that although check errors do commute with plaquette stabilizer operators,
we can detect check errors because we measure plaquette operators by two rounds of measurements,
one of which anticommutes with the error.

If measurement outcomes can be incorrectly recorded (measurement error),
we have to consider an array of syndrome bits in $1+1$-dimensional spacetime.
Let us consider faults in the $ZZ$ measurement outcomes.
One rung with a $ZZ$ check is used in two neighboring syndrome bits which have distinct time coordinates.
So, if that $ZZ$ measurement outcome was incorrect, then we would read the two syndrome bits flipped.
If our error model only allows $ZZ$ measurement outcome faults
and check errors along the bottom leg of the ladder,
then any fault pattern gives a $\ZZ_2$-chain 
whose end points ($\ZZ_2$-boundary) are nontrivial syndrome bits.
Note that a $ZZ$ measurement error near the end point of a long check error chain
does not change the fact that there is an odd number of nontrivial syndrome bits 
near the end of the chain.
Again, at low noise a matching algorithm or other decoder for the {\it two dimensional} toric code will decode the errors correctly.
We can imagine two different scenarios to understand fault tolerance.  In the simplest, check errors exist for a certain number of rounds, and then later the noise is turned off.  In this case, if decoder finds a matching in the correct homology class, then no logical error occurs.  In the other scenario, we imagine errors occurring all rounds, and then one wishes to read out the outer logical operator at some given round $r_0$.  In this case, the decoder will match errors at early rounds, but a low density of errors near round $r_0$ may remain.  However, since the outer logical operator can be read at out at $L$ different positions, where $L$ is the length of the chain.  By a majority decoding of the $L$ different outcomes, we can still decode in this setting.

\section{Fault-tolerance of the honeycomb code on torus}
\label{sec:ec}

The group of all plaquette stabilizers is a static subgroup 
of the instantaneous stabilizer groups for all time steps $r \ge 3$.
Hence, it is natural to record all the eigenvalues of the plaquette stabilizers.
We declare that these are our syndrome bits.
They are always trivial ($0 \in \FF_2$) in the absence of any faults.
In each round, we obtain a fresh subset of syndrome bits associated with exactly one-third of all plaquettes.
We show in this section that these syndrome bits are sufficient for fault tolerance,
and that the standard matching algorithm can be used in a decoding algorithm.

\subsection{Lattice of syndrome bits in spacetime}

After measuring checks of type~$r$, we infer plaquette stabilizers of type $r+1 \bmod 3$.
The plaquettes of a given type form a triangular lattice (see \cref{fig:lat})
with basis vectors 
$\mathbf t_1 = (2\cos \tfrac \pi 6, 0) \in \RR^2$ 
and $\mathbf t_2 = (\cos \tfrac \pi 6, 1+\sin \tfrac \pi 6) \in \RR^2$.
The triangular lattice of type~$r+1$ plaquettes is shifted relative to that of type~$r$
by $(0,1) \in \RR^2$.
Taking the time coordinate into account,
we see that the following three basis vectors span the lattice of syndrome bits in spacetime.
\begin{align}
\begin{cases}
	\mathbf t_1 = (2\cos \tfrac \pi 6, 0, 0)\\
	\mathbf t_2 = (\cos \tfrac \pi 6, 1 + \sin \tfrac \pi 6, 0)\\
	\mathbf t_3 = (0,1,1)
\end{cases}
\end{align}
An easier-looking basis for this lattice in $\RR^3$ is the following.
\begin{align}
\begin{cases}
	\mathbf s_1 = (\cos (\tfrac {\pi}{2} + \tfrac {0\pi}{3}), \sin (\tfrac {\pi}{2} + \tfrac {0\pi}{3}), 1) &= \mathbf t_3 \\
	\mathbf s_2 = (\cos (\tfrac {\pi}{2} + \tfrac {2\pi}{3}), \sin (\tfrac {\pi}{2} + \tfrac {2\pi}{3}), 1) &= -\mathbf t_2 + \mathbf t_3\\
	\mathbf s_3 = (\cos (\tfrac {\pi}{2} - \tfrac {2\pi}{3}), \sin (\tfrac {\pi}{2} - \tfrac {2\pi}{3}), 1) &= \mathbf t_1 - \mathbf t_2 + \mathbf t_3 
\end{cases}	
\end{align}
These basis vectors look like the three edges at the origin of the simple cubic lattice
after a shear transform, where the $(111)$ direction of the simple cubic lattice is our time direction.
Below we will actually bipartition this lattice into two congruent graphs,
each of which is isomorphic to the graph (1-skeleton) of a simple cubic lattice,
on which one can run a fault-tolerant decoding algorithm.

\subsection{Simplifying error model}

We consider a simplified error model, with perfect measurement of check operators and Pauli errors occurring randomly and independently on qubits in between the measurement of check operators.
Note that a positive threshold in this simplified error model implies a positive threshold 
even with imperfect measurement of check operators, 
as a single measurement error is equivalent to a pair of qubit Pauli errors.
Consider measuring for example some $XX$ check.
If an error of type $Y$ or $Z$ occurs on one of the qubits before measurement, 
and the same error occurs immediately after measurement, 
then the net effect is same as that of a measurement whose outcome is flipped.
In practice, however, since independent measurement errors then are equivalent to correlated qubit errors, 
a better logical error rate and a better threshold may be achieved by a decoding algorithm 
that takes this into account; 
we do not consider optimizing the decoding algorithm here.

We can further simplify the error model.
Consider some Pauli error on a qubit.  
Immediately before the error, we measure a check supported on that qubit, 
involving some Pauli operator $P_1\in \{X,Y,Z\}$ on that qubit.
Immediately after, we measure some other check that acts on that qubit, 
involving some other Pauli operator $P_2$ on that qubit.
We use these two Pauli operators as a basis for Pauli errors.
For example, if immediately before we measured an $XX$ check and immediately after we measured a $YY$ check, 
we expand errors on that qubit in the basis $X,Y$.
We use an error model where errors of types $P_1,P_2$ occur independently 
so that an error of type $P_1 P_2$ can then occur as a combination of two errors; 
of course, in practice, if the actual error model has all three error types $P_1, P_2, P_1 P_2$ equally likely, 
a better fault-tolerance performance may be achieved by an algorithm that takes this into account.
Now, the $P_2$ error can be commuted through the subsequent check;
in the example, the $Y$ error can be commuted through the $YY$ check.
Commuting through in this way does not change the state or spacetime history of measurement outcomes.
So, we will do this commutation of any $P_2$ error.

This leaves a restricted error model: 
Pauli errors may occur on a qubit of a type corresponding to whatever check was measured previously: 
if an $XX,YY,ZZ$ check is measured, then subsequently a Pauli error may occur of type $X,Y,Z$ respectively, 
with errors occurring independently.
Since there are $6 = 9 \cdot 2 / 3$ qubits per spatial unit cell
(each of $9$ types of checks has $2$ qubits, each of which is shared by $3$ checks),
the number of independent elementary faults per unit spacetime volume in this simplified error model is only $18$.

This simplified error model has an interesting relation to the fact that 
the outer operators $Q(r),Q(r+3)$ of \cref{fig:dynsuper} differ by an inner operator.
Equivalently, an inner operator equals the product of two outer operators at different rounds.
The reader may verify that if we take a partial implementation of that inner operator at some round, 
write it as a product of Paulis, and then use the rule above to commute certain Paulis to a subsequent round, 
the result indeed is two outer operators at different rounds.
Further, this should be unsurprising by considering a partially implemented inner operator with two endpoints.
At either endpoint, the inner operator causes the inferred stabilizer to flip twice.
However, a partial implementation of the outer operator $Q(r)$ causes a single flip in the inferred stabilizer.
So, two partially implemented outer operators produce the same faults.

\subsection{Decoding in the bulk}

Suppose that a measurement round $r = 0 \bmod 3$ is just done
and a single-qubit error $E = X$ occurs in the simplified error model.
At round $r+1$, exactly one type~1 check is flipped.
So, the fresh type~2 plaquette stabilizer that is inferred by the outcomes at round $r$ and $r+1$ is flipped.
If there are no other errors, this is the first time this type~2 plaquette stabilizer is flipped.
At round $r+2$, exactly one type~2 check is flipped,
but the fresh type~0 plaquette stabilizer is not flipped 
because there are two checks that are flipped in its constituents.
At round $r+3$, no check is flipped, but one type~1 plaquette stabilizer is flipped 
because of the flip of the check at round $r+2$.
So, the error $E$ causes two syndrome bit \emph{changes}, one of type~2 that is inferred right after round~$r$, 
and the other of type~1 right after round $r+2$.
These two syndrome bit changes are nearest neighbors in space, but second nearest in time.
Let us give obvious coordinates to each location of potential syndrome bit \emph{change}:
if a syndrome bit at $(p,t)$ is different from that on $(p,t+1)$,
then we say that this change occurs at $(p,t+1)$.
The two syndrome bit changes caused by $E$ are separated in spacetime by $\mathbf s_1 + \mathbf s_2$.

Since the honeycomb code is symmetric under $2\pi/3$ rotations about the time axis and 
since we are considering the time translation invariant setting,
we see that a pair of syndrome bit changes is a possible configuration
if the pair is separated by an integer combination of three vectors 
\begin{align}
\mathbf s_1 + \mathbf s_2, \quad \mathbf s_2 + \mathbf s_3, \quad \mathbf s_3 + \mathbf s_1. \label{eq:ss}
\end{align}
Since these include all generators of the error model,
only such pairs of syndrome bit changes may appear.

Hence, we have identified a decoding graph for the honeycomb code:
two copies of the Cayley graph of the free abelian group with three generators (up to boundary conditions).
There are two copies because the integer span of vectors in \eqref{eq:ss} 
only covers the even sublattice of the lattice of all potential syndrome bit changes.
Each copy Cayley graph is nothing but the 1-skeleton of a simple cubic lattice,
but in view of our lattice of all syndrome bits, this cubic lattice is shear transformed
and the~$(111)$ direction of the cubic lattice is our time direction.

Any fault, including all possible Pauli and measurement outcome errors at any location,
can be expressed in our simplified error model,
which gives a $1$-chain with $\ZZ_2$ coefficients in the decoding graph.
The inferred syndrome bit changes are represented by a $0$-chain on the decoding graph,
and the usual graph matching algorithm will identify the error $1$-chain up to plaquettes, a $2$-chain.
A Peierls argument shows that there is a positive threshold so that up to this threshold a minimum weight matching algorithm 
will match errors leaving all loops of residual errors small.

\begin{figure}
\centering
\includegraphics[width=0.9\textwidth, trim={0mm 110mm 30mm 0mm}, clip]{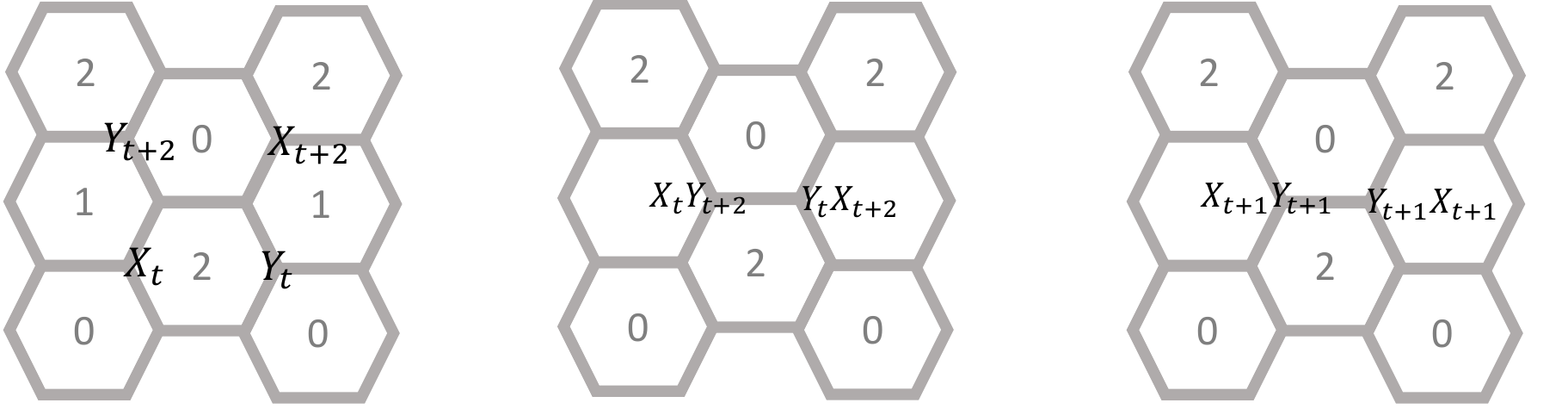}
\caption{
	A configuration of errors in spacetime that is inconsequential.
	The subscripts denote time.
	It is assumed that measurement round $r = 0 \bmod 3$ is just done.
	The three figures show equivalent errors.
	The errors in the first and second figures are equivalent 
	because they differ by an element in the ISG at the respective rounds.
	From the second figure to the third, we commute the errors through checks;
	$X_t Y_t$ commutes with the check $ZZ$ at round~$r+1$,
	and each of $Y_{t+2}$ and $X_{t+2}$ commutes with the check at round $r+2$.
	The operator in the third figure is an element of the ISG at round~$r+1$.
}
\label{fig:inconsequential}
\end{figure}

An error that forms a null-homologous cycle in the decoding graph is a product of
errors that form squares in the decoding graph.
It is routine to confirm that a square (a four-edge cycle) in the decoding graph 
corresponds to the error configuration in \cref{fig:inconsequential}
or the rotation-symmetric versions thereof.
Therefore, any null-homologous cycle in the decoding graph is inconsequential to the logical operators at later times,
and the matching decoding algorithm achieves a positive fault-tolerance threshold.%
\footnote{
One might still want to see how a partial implementation of the inner logical operator is detected
when an inner logical operator may terminate in a way that it commutes with all plaquette stabilizers.
It should be noted that plaquette stabilizers are not directly measured in the honeycomb code,
but only inferred from check measurement outcomes at different times.
So, even though the plaquette stabilizer commutes with an error,
if the error occurs in between measurement rounds, 
it can still flip the inferred syndrome bit.
This phenomenon happens also in the ladder code.
}

\subsection{Measurement of logical qubits and initialization}

Even though we have proved the fault tolerance with noisy measurements in the honeycomb code,
the discussion there assumed indefinite duration of the code dynamics.
This is not operationally sensible, 
but we can introduce meaningful time boundary conditions --- logical initialization and destructive measurements.

Let us first consider logical measurements in the $X$ and $Z$ basis, just as in the toric code.
Recall that in the toric code, to measure the logical qubit in $X$ basis,
one measure each data qubits in the $X$ basis.
These individual measurement outcomes are then postprocessed to give $X$ stabilizer measurement outcomes.
Any fault in the measurement outcome can be treated as a Pauli $Z$ error on a data qubit,
and this means that without loss of fault tolerance we may assume that the final single-qubit measurements are noiseless.
$X$ errors do not matter since they commute with the measurement,
and the reconstructed syndrome enables us to correct $Z$ errors.
Once we correct $Z$ errors, we choose any representative of the $X$ logical operator,
and the logical measurement outcome is then inferred fault-tolerantly.

Since the honeycomb code at any moment is equivalent (by a depth~$1$ quantum circuit) to the toric code state,
a similar strategy can be used to measure \emph{outer} logical operators.
Suppose we just have completed round~$r = 0 \bmod 3$.
The two qubits on any type~$0$ edge form an effective qubit,
and there are operators, commuting with the check, corresponding to the single-qubit $X$ and $Z$ of the superlattice toric code.
These are easy to read off from the \cref{fig:toric} by the rule that type~$0$ plaquette stabilizers must
correspond to the $Z$-type plaquette terms of the superlattice toric code,
and type~$1$ and~$2$ plaquette stabilizers to the $X$-type terms.
For example, on a type $0Y$ edge, a $Y$ operator on one of the two qubits corresponds to $Z$ of the superlattice toric code.
The operator corresponding to a single-qubit $X$ of the superlattice toric code,
has weight~$2$.
Then, we can simply measure these effective-one-qubit operator on each edge.
The reason that this is fault tolerant is the same as that for the toric code.

In practice, it may be better to insist on single-qubit measurements,
since the code state is anyway destroyed and we can infer the outcome of any two-qubit measurement by one-qubit measurements.
This way, the architectural requirement is to have only one type of two-qubit Pauli measurement per direction
and arbitrary one-qubit Pauli measurement per data qubit.

Initialization can be done in a similar manner.
In the usual toric code, initialization, for example, in the $X$ basis 
starts by preparing each data qubit in the $X$ basis independently,
and then we project the state into the code space by syndrome measurements.
In our honeycomb code, we can prepare each pair of qubits on an edge in a particular state independently
specified by the effective-one-qubit operators,
and start our check measurement sequence.

\section{Boundary Conditions}
It may be desirable to have a code which can be realized with a planar geometry, rather than a torus.
In this case, we need to be able to introduce boundary conditions.
Before considering how to introduce boundary conditions using a sequence of measurements of pairwise operators,
let us first consider how to modify the bulk ISG near a boundary to introduce boundary conditions.

\begin{figure}[b]
  \centering
  \includegraphics[width=0.5\textwidth]{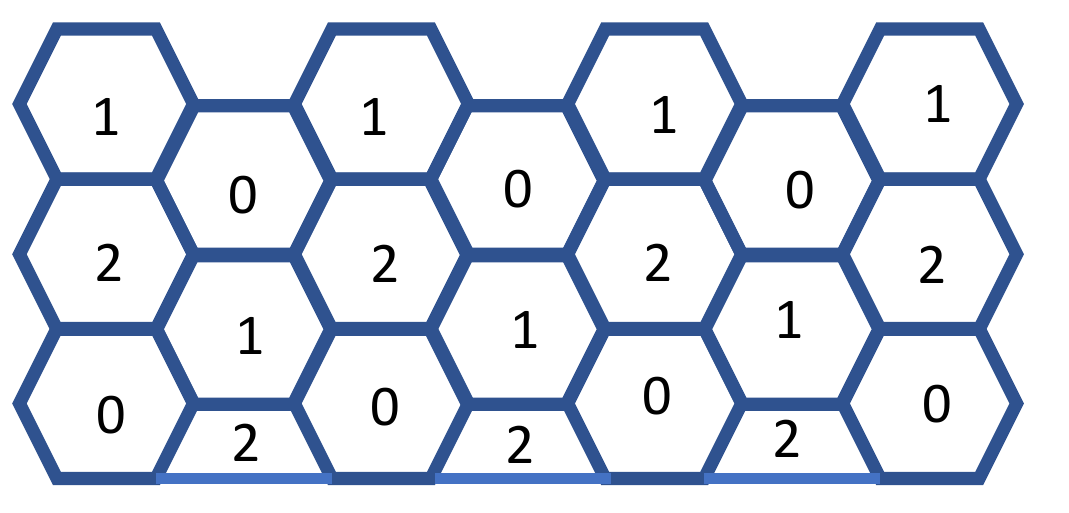}
  \caption{Introducing a boundary to ISG.  }
  \label{figarmchair}
\end{figure}

Consider \cref{figarmchair}.  We will only consider introducing a boundary at the bottom of the figure.  The left and right edge will be extended as desired and then joined with periodic boundary conditions, giving an annulus.  Another boundary will be added at the top, possibly first extending the figure upwards, of course.  One may call this kind of boundary an ``armchair'' boundary, similar to the use of the term in carbon nanotubes.

We have added some additional edges to the bottom of the figure so that all vertices are trivalent.  This creates also some square plaquettes.  These added edges are type $0$ and the added plaquettes are type $2$: the type of an edge depends on the plaquettes at its ends, regardless of whether those plaquettes are squares or hexagons.

The checks corresponding to these added edges will also be pairwise, being a product of two Pauli operators.  
The particular Pauli operators chosen for these checks will be such that for every qubit, the three checks incident involve distinct Pauli operators on that qubit.  Indeed, for all figures that follow, we assume that the Pauli operators are chosen in this way.

Suppose we take the ISG defined above for the honeycomb code after any given round $r\geq 3$, restricting to just the terms in the ISG which are supported on the lattice with boundary.
Then add the following additional stabilizer generators.  For every square plaquette, add the product of checks around that plaquette as a plaquette operator.  Also, if $r=2\mod 3$, add the checks supported on the added type $2$ edges.  Call the resulting group the ``boundary ISG.''

One may verify that these added generators give a gapped boundary.  Indeed, using a similar construction to \cref{fig:toric}, the boundary conditions are a so-called ``smooth'' boundary for $r=1 \mod 3$ and a so-called ``rough'' boundary for $r=0,2 \mod 3$.  These different smooth and rough boundary conditions correspond to what are also called electric and magnetic boundary conditions.

We can immediately guess then that if we start with the boundary ISG after round $r$ and then measure checks of type $r+1$, there is no problem if $r=0,1\mod 3$, but that there will be a problem if $r=2 \mod 3$.
The reason is, going from $r=0 \mod 3$ to $r=1 \mod 3$ or $r=1 \mod 3$ to $r=2 \mod 3$ interchanges electric and magnetic (or smooth and rough) boundary conditions, which matches what we expect since the outer logical operator changes from an electric to a magnetic string and vice-versa after each measurement round.  However, going from $r=2\mod 3$ to $r=0 \mod 3$ leaves the boundary conditions of the same type.

We can see that there is a problem for $r=2 \mod 3$ in a different way also: measuring all type $2$ checks and then all type $0$ checks will measure the inner logical operator which is the product of all checks on the bottom boundary, i.e., this is the product of checks on the bottom of the squares times those on the bottom of the type $0$ hexagons at the bottom of the figure.

One simple remedy is as follows.  Consider a system on an annulus of length $L$.  Start with an ISG after measuring type 0 checks.  Pick a strip of size $O(1)$ extending from top to bottom boundary.  In this strip measure checks $0,1,2,0$ in sequence.  This allows one to measure all of the plaquette stabilizers (since every plaquette of type $r+1 \mod 3$ can be measured by measuring checks $r,r-1 \mod 3$ successively).  It does not however lead to a measurement of the inner logical operator on the boundary since the strip does not extend the full length of the boundary.  
In fact, we may do this on of order $L$ nonoverlapping strips in parallel.
Then, choose another set of order $L$ nonoverlapping strips, and do the same sequence, repeating until all plaquettes are measured.

One may wonder: is it possible to retain the periodic sequence of measuring $0,1,2$ checks cyclically in the bulk?  For example, could some clever choice of pairwise checks on the boundary (perhaps measuring the checks with period $6$ or higher near the boundary) be found to avoid leaking information?
In fact there is a topological obstruction to doing this using only pairwise checks, at least in a certain sense that we explain below.

We use the Majorana language throughout.
Our goal is to analyze the system on an annulus.  Consider some arbitrary trivalent graph, with qubits on the vertices.  Give the vertices two coordinates $x,y$ in the plane, with the system periodic in the $x$ direction with period $L$.  However, to introduce some ideas we consider the system where the $x$ coordinate is {\it not} periodic; instead we consider an infinite system in the $x$ direction.  In this case, we can introduce a topological invariant.  We assume from here on that after any number of measurements (perhaps after some but not all checks have been measured in a round) that the ISG includes plaquette stabilizers guaranteeing (in the Majorana language) that the product of gauge fields around any homologically trivial loop is $+1$.  Further, assume that the ISG is generated by these plaquette stabilizers as well as by operators of form $\gamma^0_j \gamma^0_k$ times a product of gauge fields on a path from $j$ to $k$.
Note that a pairwise check is such a $\gamma^0_j \gamma^0_k$ times such a product of gauge fields when $j,k$ are neighbors.

In this case, we can easily see how the ISG changes when checks are measured.
Indeed, with open boundary conditions, since the system is topologically trivial, there is no need to specify the path; with periodic boundary conditions, we need to specify the homology class of the path.  So, we can represent the ISG by drawing a set of ``dimers.''  Each generator $\gamma^0_j \gamma^0_k$ (times the product of gauge fields) is represented by 
a ``dimer,'' which simply means an unordered set of two elements $\{j,k\}$.  The dimer can be represented pictorially by drawing a line from $j$ to $k$ (with the line not necessarily on the edges of the graph).
Further, every site will be in exactly one dimer so that the ISG has the correct number of generators.

Then, the effect of measuring a check on a pair $k,l$ is as follows: if there are dimers $\{i,k\}$ and $\{j,l\}$, then after measuring the check we have dimers $\{i,j\}$ and $\{k,l\}$.
The reader may then see what happens when measuring on a hexagon.  Label the sites $1,2,3,4,5,6$.  Start with an ISG where 3 edges of a hexagon contain dimers, say $\{1,2\},\{3,4\},\{5,6\}$ and then measure checks on the other three edges.  The dimers change to $\{1,4\},\{2,3\},\{5,6\}$, then $\{1,6\},\{2,3\},\{4,5\}$ as we measure checks on edges $2,3$ and $4,5$ in turn.  One may pictorially think of this as one of the dimers (in this case, the $\{1,2\}$ dimer) splitting in two, with one of the two halves ``winding around'' the hexagon before rejoining its partner.  The final measurement of check $5,6$ then measures the product of gauge fields around the hexagon.

This dynamics for dimers has an invariant: the number of dimers connecting sites with $x$ coordinate $>0$ to those with $x$ coordinate $<0$ is invariant modulo $2$.  Indeed, $0$ can be replaced by any number here.  We can see this pictorially as counting the number of dimer lines which are cut by a line drawn at $0$.  Much of our construction is inspired by \cite{Kivelson_1988}.

Remark: in fact, our graph has a two sublattice structure (i.e., there is a perfect cut into two sets $A,B$), because we wish all plaquettes to have even length.  Using this structure, we can promote this $\ZZ_2$ invariant into an integer invariant by ordering the sites in the dimer from $A$ to $B$, and counting the number with a sign depending on whether site $A$ is $<0$ and site $B$ is $>0$ or vice-versa.  However, we will not need to consider this in what follows.

Now we consider the case of an annulus.  In this case, everything is the same, except that we need to specify the homology class of the path from $j$ to $k$ for each dimer $\{j,k\}$.

Then, the effect of measuring a check on a pair $k,l$  connected by some path $P_{k,l}$ is as follows: if there are dimers $\{i,k\}$ and $\{j,l\}$ connected by paths $P_{i,k}$ and $_{j,l}$, then after measuring the check we have one dimer $\{i,j\}$ connected by a path $P_{i,k}+P_{j,l}+P_{k,l}$ and another dimer $\{k,l\}$ connected by a path $P_{k,l}$.  Here, the sum of paths means the sum of $\ZZ_2$ chains.
Specifying the homology class of the path allows us to continue to define this invariant for a periodic system: count, modulo $2$, the sum over dimers of the number of times the path for that dimer crosses some fixed line from one edge of the annulus to the other.  That is, treat the sum of paths as a chain and compute its $\ZZ_2$ intersection with some other chain.

Remark: this kind of invariant for periodic systems is long-studied in the condensed matter physics literature~\cite{Kivelson_1988}, but typically rather than explicitly specifying the path, one considers dimers whose endpoints are nearby and then one implicitly uses a shortest path.

Now, suppose one has found some clever sequence of checks at the boundaries so that the inner logical operator of the code is not measured.  Consider an annulus, measuring cyclically $0,1,2$ in the bulk, and use this ``clever sequence'' at the top boundary but continue to use a ``naive sequence'' at the bottom boundary, where the naive sequence is simply to use the armchair boundary conditions and measure checks $0,1,2$ cyclically at the bottom boundary also.

Start with an ISG after round $0$ where are all dimers are nearest neighbors on the graph, and all paths are shortest paths.
Then measure checks $1,2,0$ in sequence.  Then, in the bulk, the dimers return to their initial state.  However, at the bottom (naive) boundary, one finds that the sum of paths has changed by adding a nontrivial homology representative.  (Of course, one has also changed the ISG by adding the inner logical operator to the ISG too).
Hence, no matter what sequence is chosen at the top boundary, the sum of paths at that boundary must also add a nontrivial homology representative.  Heuristically, one may say that an odd number of Majoranas have ``wound around'' the annulus at the bottom edge, and so the same must happen at the top edge.  Then, the measurement of checks reveals the inner logical operator also at the 
top boundary!

This topological obstruction does not, however, prevent transitions which are cyclic in the bulk but which use more complicated, nonpairwise checks on the boundary (which in turns can be built out of pairwise checks with ancillas).  For example, one can follow a sequence $0,1,2$, finishing at $2$ with rough boundary conditions.  Then, before measuring $0$, one can shrink the code by single qubit measurements near the boundary so that $0$ becomes smooth (while $2$ stays rough), and then measure $0$.  One may continue in this fashion, shrinking as necessary, and then uses nonpairwise checks to grow the code to compensate the shrinking.

\bibliographystyle{apsrev4-1}
\nocite{apsrev41Control}
\bibliography{refs}

\appendix
\section{A Toy Model of Dynamical Quantum Memory: Not an Error Correcting Code}
\label{app}
In certain models considered in the literature, one intersperses fast scrambling unitary dynamics with measurements.  The interesting thing about these models is that in many cases there is a phase transition depending on the measurement rate.   With frequent measurements, the density matrix of the system rapidly purifies even if started in a maximally mixed state.  On the other hand, with less frequent measurements, if the system is started in a mixed state it can remain mixed for very long times.  Related to this is that if the mixed state on the system results from entanglement with a reference, then some entanglement can be preserved for a long time.

However, such fast scrambling dynamics typically will {\it not} lead to useful protection against external noise.  Let us consider this in a very simple toy model, studied in \cite{Fidkowski_2021}.

Suppose the model is a system of qubits, and one has a dynamics which alternates two different steps.  First, measure a single qubit in the $Z$-basis.  Then, apply a random Clifford to the entire system.  Up to conjugation by unitaries, this is equivalent to measuring a product of Pauli operators at each step, with the product chosen uniformly at random from all possibilities (other than the identity operator).

Suppose we start with maximally mixed system on $N$ qubits.  After any number of steps, there is some number, $K$, of stabilizers.  Then, if we assume for simplicity that the next measurement is chosen uniformly from all products of Paulis (including the identity), the probability that the next measurement commutes with all of the stabilizers is $2^{-K}$.  If the measurement does not commute, then the number of stabilizers remains unchanged after measurement, while if it does commute, then the number of stabilizers may increase by $1$ if the measured operator is not in the stabilizer group.  The probability that it is in the stabilizer group is $4^{-(N-K)}$, so it is indeed likely that the measured operator will not be in the group.

Hence, the time required for the state to become pure (so that $K=N$) is exponential in $N$.
Similarly, if the initial state is a maximally entangled pure state between the system and reference, the time for all entanglement between system and reference to be lost is exponential in $N$ (note here that there is no external noise, so the state remains pure for all time, so if the stabilizer group cannot be generated by independent stabilizer subgroups on system and reference then there is entanglement).

By the same token, however, since the probability that the operator measured is in the stabilizer group is exponentially small in $N$ (for any $K$), there is only an exponentially small probability to detect if external noise has been applied.  Suppose some noise in the form of a  random Clifford operator is applied to the system.  This may change the entangelement between system and reference and change the stabilizers.  However (except for exponentially small probability) each measurement outcome without noise is $+1$ with probability $1/2$ and $-1$ with probability $1/2$ and we have the same distribution of measurement outcomes with this noise applied. Thus, this is not a useful quantum error correcting code.
\end{document}